\documentclass[copyright,creativecommons,sharealike]{eptcs}

\usepackage{tikz}
\usetikzlibrary{arrows,backgrounds,decorations,decorations.pathmorphing,positioning,fit,automata,shapes,snakes}
\usepackage{amsmath, amssymb, amsfonts}
\usepackage{color,graphicx, enumerate, subfig}
\usepackage[ruled,vlined,linesnumbered]{algorithm2e}
\usepackage{url}
\usepackage{balance}
\usepackage{multirow,wrapfig}
\usepackage[strings]{underscore}
\newcommand{\ie}{{i.e.}\xspace}

\newcommand{\eg}{{e.g.}\xspace}
\newcommand{\etc}{{etc.}\xspace}

\newcommand{\secref}[1]{Sec.~\ref{sec:#1}}

\newcommand{\figref}[1]{Fig.~\ref{fig:#1}}

\newcommand{\ignore}[1]{}

\newtheorem{thm}{Theorem}

\newcommand{\assign}{{\tt :=}}

\newcommand{\true}{{\tt T}}
\newcommand{\false}{{\tt F}}

\newcommand{\ca}[1]{\mathbb{{#1}}}
\newcommand{\Type}{\bL^{\mathbb{S}}}
\newcommand{\vCTL}{${\bf L}$CTL\xspace}
\newcommand{\val}{val}
\newcommand{\vinit}{\upsilon_{init}}
\newcommand{\tf}[1]{\textsf{{#1}}}

\newcommand{\lr}{\leftrightarrow}

\newcommand{\bL}{{\bf L}}

\newcommand{\lk}[1]{{\tt lock(}{#1}{\tt )}}

\newcommand{\wait}[2]{{\tt wait(}{#1},{#2}{\tt )}}
\newcommand{\sig}[1]{{\tt signal(}{#1}{\tt )}}

\title{Towards Algorithmic Synthesis of Synchronization for Shared-Memory Concurrent Programs}
\author{Roopsha Samanta
\institute{Computer Engineering Research Centre, 
\\ The University of Texas at Austin.} \\
\email{roopsha@cs.utexas.edu}}
\def\titlerunning{Algorithmic Synthesis of Synchronization}
\def\authorrunning{R. Samanta}
\begin{document}
\maketitle

\pagestyle{plain}

\begin{abstract}
We present a framework that takes a concurrent program composed of
unsynchronized processes, along with a temporal specification of their
global concurrent behaviour, and {\em automatically} generates a
concurrent program with synchronization ensuring {\em correct} global
behaviour. Our methodology supports finite-state concurrent programs
composed of processes that may have local and shared variables, may be
straight-line or branching programs, may be ongoing or terminating,
and may have program-initialized or user-initialized variables.  The
specification language is an extension of propositional Computation
Tree Logic (CTL) that enables easy specification of safety and
liveness properties over control and data variables. The framework
also supports synthesis of synchronization at different levels of
abstraction and granularity.

\end{abstract}
\section{Introduction}\label{sec:intro}

Shared-memory concurrent programs are ubiquitous in today's era of
multi-core processors. Unfortunately, these programs are hard to write
and even harder to verify. We assert that one can simplify the
design and analysis of (shared-memory) concurrent programs by, first,
manually writing synchronization-free concurrent programs, followed
by, automatically synthesizing the synchronization code necessary for
ensuring the programs' correct concurrent behaviour.  This particular
approach to synthesis of concurrent programs was first developed in
\cite{EmeCla82, AttEme89} and was revisited more recently in
\cite{JanMyc06,VYY09,VYY10}. The early synthesis papers focused on
propositional temporal logic specifications and restricted models of
concurrent programs such as synchronization skeletons. Even when
dealing with finite-state programs, it is highly cumbersome to express
properties over functions and predicates of program variables using
propositional temporal logic. Besides, synchronization skeletons that
suppress data variables and computations are often inadequate
abstractions of real-world concurrent programs.  The more recent
synthesis approaches have fairly sophisticated program models.
However, they are applicable for restricted classes of specifications
such as safety properties, and entail some possibly restrictive
assumptions. For instance, it is almost always assumed that all data
variables are initialized within the program to specific values,
thereby disallowing any kind of user or environment input to a
concurrent program.  The presence of local data variables is also
rarely accounted for or treated explicitly.  Finally, there has been
limited effort in developing adaptable synthesis frameworks that are
capable of generating synchronization at different levels of
abstraction and granularity.

In this paper, we present a comprehensive treatment of synthesis of
synchronization for concurrent programs with CTL-like specifications
over program variables. We support finite-state concurrent programs
composed of processes that may have local and shared variables, may be
straight-line or branching, may be ongoing or terminating, and may be
executed as a closed system (with no external environment) or with an
external environment that may initialize the values of the program
variables or read the values of the program variables at any point in
the programs' execution. We propose an extension to propositional CTL
that helps express properties over program locations and data
variables. These properties may be syntactic, \eg, $\tf{A}\tf{G} \,
\neg(loc_1 = l_1 \, \wedge \, loc_2 = l_2)$, specifying that the first
and the second process cannot simultaneously be in locations $l_1$ and
$l_2$, respectively, or semantic, \eg, $\tf{A}\tf{G} \, (v_1 =
\upsilon \, \Rightarrow \, \tf{A}\tf{F} (v_2 = \upsilon + 1))$,
specifying that if the value of variable $v_1$ is $\upsilon$, then it
is inevitable that the value of variable $v_2$ be $\upsilon+1$, or
both syntactic and semantic.  Furthermore, as is evident from the
above examples, these properties may express safety as well as
liveness requirements. Finally, we support the synthesis of
synchronization in the form of conditional critical regions (CCRs), or
based on lower-level synchronization primitives such as locks and
condition variables. In the latter case, the synthesized
synchronization can be either coarse-grained or fine-grained. 

Given a concurrent program $P$ composed of synchronization-free
processes, $P_1, P_2, \ldots, P_k$, and a temporal logic specification
$\phi_{spec}$ specifying the expected concurrent behaviour, the goal
is to obtain synchronized processes, $P^s_1, P^s_2, \ldots, P^s_k$,
such that the concurrent program $P^s$ resulting from their {\em
asynchronous composition} satisfies $\phi_{spec}$. This is effected in
several steps in our proposed approach. The first step involves
specifying the concurrency and operational semantics of the
unsynchronized processes as a temporal logic formula $\phi_P$. We help
mitigate the user's burden of specification-writing by automatically
generating $\phi_P$.  The second step involves construction of a
tableau $T_{\phi}$, for $\phi$ given by $\phi_P \wedge \phi_{spec}$.
If the overall specification is found to be satisfiable, the tableau
yields a global model $M$, based on $P_1, P_2, \ldots, P_k$ such that
$M \models \phi$. The next step entails decompositon of $M$ into the
desired synchronized processes $P^s_1, \ldots, P^s_k$ with
synchronization in the form of $CCRs$. The last step comprises a
mechanical compilation of the synthesized CCRs into both
coarse-grained and fine-grained synchronization code based on 
locks and condition variables. 

To construct the tableau $T_{\phi}$, we adapt the tableau-construction
for propositional CTL to our extended specification language over
variables, functions and predicates. When there exist
environment-initialized variables, we present an initial brute-force
solution for modifying the basic approach to ensure that $P^s$
satisfies $\phi_{spec}$ for all possible initial values of such
variables.  Also, we address the effect of local variables on the
permitted behaviours in $P^s$ due to limited observability of global
states, and discuss solutions. 

The paper is structured as follows. We begin by introducing our
specification language and program model in \secref{prelim}.  We
present a basic algorithmic framework in \secref{approach}, focussing
on the formulation of $\phi_P$, tableau construction, model generation
and extraction of CCRs. We then address extensions of the basic
framework to deal with uninitialized variables, local variables,
different synchronization primitives and multiple processes 
in \secref{ext}. We conclude with a discussion of related and future
work in \secref{disc}.

\section{Formal Framework}\label{sec:prelim}
\subsection{A vocabulary ${\bf L}$}\label{sec:alpha}

\noindent{\em Symbols of ${\bf L}$}: We fix a vocabulary ${\bf L}$
that includes a set $\bL^{\ca{V}}$ of variable symbols (denoted $v$,
$v_1$ \etc), a set $\bL^{\ca{F}}$ of function symbols (denoted $f$,
$f_1$ \etc), a set $\bL^{\ca{B}}$ of predicate symbols (denoted $B$,
$B_1$ \etc), and a non-empty set $\Type$ of {\em sorts}.  $\Type$
contains the special sort ${\tt bool}$,  along with the special sort
{\tt location}. Each variable $v$ has associated with it a sort in
$\Type$, denoted $sort(v)$. Each function symbol $f$ has an associated
arity and a sort: $sort(f)$ for an $m$-ary function symbol is an $m +
1$-tuple $<\sigma_1,\ldots, \sigma_m, \sigma>$ of sorts in $\Type$,
specifying the sorts of both the domain and range of $f$.  Each
predicate symbol $B$ also has an associated arity and sort: $sort(B)$
for an $m$-ary predicate symbol is an $m$-tuple $<\sigma_1,\ldots,
\sigma_m>$ of sorts in $\Type$.  Constant symbols (denoted $c$, $c_1$
\etc) are identified as the $0$-ary function symbols, with each
constant symbol $c$ associated with a sort, denoted $sort(c)$, in
$\Type$.  The vocabulary ${\bf L}$ also explicitly includes the
distinguished equality predicate symbol $=$, used for comparing
elements of the same sort. \\

\noindent{\em Syntax of ${\bf L}$-terms and ${\bf L}$-atoms}: 
Given any set of variables $V \subseteq \bL^{\ca{V}}$, we inductively
construct the set of $\bL$-terms and $\bL$-atoms over $V$, 
using sorted symbols, as follows:

\begin{itemize}
\item Every variable of sort $\sigma$ is a term of sort $\sigma$.
\item If $f$ is a function symbol of sort
$<\sigma_1,\ldots,\sigma_m,\sigma>$, and $t_j$ is a term of sort 
$\sigma_j$ for $j \in [1,m]$, then $f(t_1,\ldots, t_m)$ is a 
term of sort $\sigma$.
In particular, every constant of sort $\sigma$ is a term of sort 
$\sigma$.
\item If $B$ is a predicate symbol of sort $<\sigma_1, \ldots,
\sigma_m>$, and $t_j$ is a term of sort $\sigma_j$ for $j \in
[1,m]$, then $B(t_1,\ldots, t_m)$ is an atom.
\item If $t_1$, $t_2$ are terms of the same sort, $t_1 = t_2$ 
is an atom.
\end{itemize}

\noindent{\em Semantics of ${\bf L}$-terms and  ${\bf L}$-atoms}: 
Given any set of variables $V \subseteq \bL^{\ca{V}}$, an {\em
interpretation} $I$ of symbols of $\bL$, and $\bL$-terms and $\bL$-atoms
over $V$ is a map satisfying the following: 

\begin{itemize}
\item Every sort $\sigma \in \Type$ is mapped to a nonempty domain
$D_{\sigma}$. In particular, the sort ${\tt bool}$ is mapped to the
Boolean domain $D^{\tt bool}:\{\true, \false\}$, and the sort {\tt location} is
mapped to a domain of {\em control locations} in a program. 
\item Every variable symbol $v$ of sort $\sigma$ is mapped to an
element $v^{I}$ in $D_{\sigma}$.
\item Every function symbol $f$, of sort $<\sigma_1, \ldots, \sigma_m,
\sigma>$ is mapped to a function $f^{I}: D_{\sigma_1} \times \ldots D_{\sigma_m} \to
D_{\sigma}$. In particular, every constant symbol $c$ of sort $\sigma$
is mapped to an element $c^{I} \in D_{\sigma}$. 
\item Every predicate symbol $B$ of sort $<\sigma_1 \ldots \sigma_m>$ 
is mapped to a function $D_{\sigma_1} \times \ldots D_{\sigma_m} \to
D^{\tt bool}$. 
\end{itemize}

Given an interpretation $I$ as defined above, the valuation
$\val^I[t]$ of an $\bL$-term $t$ and the valuation $\val^I[G]$ of an
$\bL$-atom $G$ are defined as follows:

\begin{itemize}
\item For a term $t$ which is a variable $v$, the valuation is $v^I$.
\item For a term $f(t_1, \ldots, t_m)$, the valuation $\val^I[f(t_1, \ldots,
t_m)] = f^I(\val^I[t_1], \ldots, \val^I[t_m])$. 
\item For an atom $G(t_1, \ldots, t_m)$, the valuation $\val^I[G(t_1, \ldots,
t_m)] = \true$ iff $G^I(\val^I[t_1], \ldots, \val^I[t_m]) = \true$. 
\item For an atom $t_1 = t_2$, $\val^I[t_1 = t_2] = \true$ iff 
$\val^I[t_1] = \val^I[t_2]$.  
\end{itemize}

In the rest of the paper, we assume that the interpretation of
constant, function and predicate symbols in $\bL$ is known and fixed.
We further assume that the interpretation of sort symbols to specific
domains is known and fixed. With some abuse of notation, we shall
denote the interpretation of all constant, function and predicate
symbols simply by the symbol name, and identify sorts with their
domains. Examples of some constant, function and predicate symbols
that may be included in $\bL$ are: constant symbols $0, 1, 2$, function
symbols $+,-$, and predicate symbols $<,>$ over the integers, function
symbols $\vee, \neg$ over ${\tt bool}$, the constant symbol $\varphi$
(empty list), function symbol $\bullet$ (appending lists) and
predicate symbol $null$ (emptiness test) over lists, \etc. Finally,
when the interpretation is obvious from the context, we denote the
valuations $\val^I[t]$, $\val^I[G]$ of terms $t$ and atoms $G$ simply as
$\val[t]$, $\val[G]$, respectively.  



\subsection{Concurrent Programs}\label{sec:prog}

In our framework, we consider a (shared-memory) concurrent program to
be an asynchronous composition of a non-empty, finite set of
processes, equipped with a finite set of program variables that range
over finite domains. We assume a simple concurrent programming
language with assignment, condition test, unconditional goto,
sequential and parallel composition, and the synchronization primitive
- conditional critical region (CCR) \cite{Hoare71,Hansen81}. A concurrent
program $P$ is written using the concurrent programming language, in
conjunction with $\bL$-terms and $\bL$-atoms. We assume that the sets
of (data and control) variables, functions and predicates available
for writing $P$ are each finite subsets of $\bL^{\ca{V}}$,
$\bL^{\ca{F}}$ and $\bL^{\ca{B}}$, respectively. 


A concurrent program is given as $P:: [{\tt declaration}] \, [P_1 {\tt
\parallel} \ldots {\tt \parallel} P_k]$, with $k > 0$. The declaration
consists of a finite sequence of declaration statements, specifying
the set of shared data variables $X$, their domains, and possibly
initializing them to specific values. For example, the declaration
statement, $v_1,v_2:\{0,1,2,3\} \; {\tt with} \; v_1 = 0$, declares two
variables $v_1$, $v_2$, each with (a finite integer) domain
$\{0,1,2,3\}$, and initializes the variable $v_1$ to the value $0$.
The initial value of any uninitialized variable is assumed to be a
user/environment input from the domain of the variable. 

A process $P_i$ consists of a declaration of local data variables
$Y_i$ (similar to the declaration of shared data variables in $P$),
and a finite sequence of labeled, {\em atomic} instructions, $l:
inst$.  We denote the unique instruction at location $l$ as $inst(l)$.
The set of data variables $Var_i$ accessible by $P_i$ is given by $X
\cup Y_i$. The set of labels or {\em locations} of $P_i$ is denoted
$L_i = \{l_i^0, \ldots, l_i^{n_i}\}$, with $l_i^0$ being a designated
start location. Unless specified otherwise\footnote{A user may define
an atomic instruction (block) as a sequence of assignment, conditional 
and goto statements}, an atomic instruction $inst$ is an assignment,
condition test, unconditional goto, or CCR. An assignment instruction
$A$, given by $(v_{i_1}, \ldots, v_{i_q}) \, {\bf \assign} \, (t_1,
\ldots, t_q)$, is a parallel assignment of $\bL$-terms $t_1, \ldots,
t_q$, over $Var_i$, to the data variables $v_{i_1}, \ldots, v_{i_q}$
in $Var_i$. Upon completion, an assignment statement at $l_i^r$
transfers control to the next location $l_i^{r+1}$. A condition test,
${\tt if} \; (G) \; l_{if},\, l_{else}$, consists of an $\bL$-atom $G$
over $Var_i$, and a pair of locations $l_{if}, l_{else}$ in $L_i$ to
transfer control to if $G$ evaluates to $\true$, $\false$,
respectively. The instruction ${\tt goto} \; l$ is a transfer of
control to location $l \in L_i$. A CCR is a guarded insruction block,
$G \to inst\_block$, where the enabling guard $G$ is an $\bL$-atom
over $Var_i$ and $inst\_block$ is a sequence of assignment,
conditional and goto statements. The guard $G$ is evaluated atomically
and if found to be $\true$, the corresponding $inst\_block$ is
executed atomically, and control is transferred to the next location.
If $G$ is found to be $\false$, the process {\em waits} at the same
location till $G$ evaluates to $\true$. An unsynchronized process does
not contain CCRs.  



We model the asynchronous composition of concurrent processes by the
nondeterministic interleaving of their atomic instructions. Hence, at
each step of the computation, some process, with an enabled
transition, is nondeterministically selected to be executed next by a
scheduler. The set of program variables is denoted $V = Loc \cup Var$,
where $Loc = \{loc_1, \ldots, loc_k\}$ is the set of control variables
and $Var = Var_1 \cup \ldots \cup Var_k$ is the set of data variables.
The semantics of the concurrent program $P$ is given by a transition
system $(S, S^0, R)$, where $S$ is a set of states, $S_0 \subseteq S$ is
a set of initial states and $R \subseteq S \times S$ is the transition relation. Each state $s
\in S$ is a valuation of the program variables in $V$. We denote
the value of variable $v$ in state $s$ as $\val^s[v]$, and the
corresponding value of a term $t$ and an atom $G$ in state $s$ as
$\val^s[t]$ and $\val^s[G]$, respectively. $\val^s[t]$ and $\val^s[G]$
are defined inductively as in \secref{alpha}. The domain of each
control variable $loc_i \in V$ is the set of locations $L_i$, and the
domain of each data variable is determined from its declaration.
The set of initial states $S_0$ corresponds to all states $s$
with $\val^s[loc_i] = l_i^0$ for all $i \in [1,k]$, and $\val^s[v] =
\vinit$, for every data variable $v$ initialized in its declaration
to some constant $\vinit$.  There exists a transition 
from state $s$ to $s'$ in $R$, with $\val^s[loc_i] =
l_i$, $val^{s'}[loc_i] = l'_i$ and $\val^{s'}[loc_j] = \val^s[loc_j]$
for all $j \neq i$, iff there exists a corresponding {\em local move} in
process $P_i$ involving instruction $inst(l_i)$, such that:
\begin{enumerate}
\item $inst(l_i)$ is the
assignment instruction: $(v_{i_1}, \ldots, v_{i_q}) \, {\bf \assign}
\, (t_1,\ldots, t_q)$, for each variable $v_{i_j}$ with $j \in [1,q]$: 
$\val^{s'}[v_{i_j}] = \val^s[t_j]$, for all other data variables $v$:
$\val^{s'}[v] = \val^s[v]$, and $l'_i$ is the next location in $P_i$
after $l_i$, or, 
\item $inst(l_i)$ is the condition test: ${\tt if} \;
(G) \; l_{if},\, l_{else}$, the valuation of all data variables in
$s'$ is the same as that in $s$, and either $\val^s[G]$ is $\true$ and
$l'_i = l_{if}$, or $\val^s[G]$ is $\false$ and $l'_i = l_{else}$, or,
\item  $inst(l_i)$ is ${\tt goto} \; l$,  the valuation of all data
variables in $s'$ is the same as that in $s$, and $l'_i = l$, or, 
\item $inst(l_i)$ is the CCR $G \to inst\_block$, $\val^s[G]$ is $\true$,
the valuation of all data variables in $s'$ correspond to the atomic
execution of $inst\_block$ from state $s$, and $l'_i$ is the next
location in $P_i$ after $l_i$.  
\end{enumerate}

\noindent We assume that $R$ is total. For terminating processes
$P_i$, we assume that $P_i$ ends with a special instruction,
$halt: {\tt goto} \; halt$.  

\subsection{Specifications}\label{sec:spec}

Our specification language, \vCTL, is an extension of propositonal
CTL, with formulas composed from ${\bf L}$-atoms.  While one can use
propositional CTL for specifying properties of finite-state programs,
\vCTL enables more natural specification of properties of concurrent
programs communicating via typed shared variables.  We describe the
syntax and semantics of this language below.  \\

\noindent{\em Syntax}: 
Given a set of variables $V \subseteq
\bL^{\ca{V}}$, we inductively construct the set of (\vCTL) {\em formulas}
over $V$, using ${\bf L}$-atoms, in conjunction with the
propositional operators $\neg, \vee$ and the temporal operators
$\textsf{A},\tf{E},\tf{X},\tf{U}$,
along with the process-indexed next-time operator $\tf{X}_i$:

\begin{itemize}
\item Every ${\bf L}$-atom over $V$ is a formula. 
\item If $\phi_1$, $\phi_2$ are formulas, then so are $\neg \phi_1$
and $\phi_1 \vee \phi_2$. 
\item If $\phi_1$, $\phi_2$ are formulas, then so are $\tf{E}\tf{X} \, \phi_1$,
$\tf{E}\tf{X}_i \, \phi_1$, $\tf{A}[\phi_1 \, \tf{U} \, \phi_2]$ and
$\tf{E}[\phi_1 \, \tf{U} \, \phi_2]$.
\end{itemize}

We use the following standard abbreviations: $\phi_1 \wedge \phi_2$
for $\neg(\neg \phi_1 \vee \neg \phi_2)$, $\phi_1 \to \phi_2$ for
$\neg \phi_1 \vee \phi_2$, $\phi_1 \lr \phi_2$ for $(\phi_1 \to
\phi_2) \wedge (\phi_2 \to \phi_1)$, $\tf{A}\tf{X} \, \phi$ for $\neg
\tf{E}\tf{X} \, \neg
\phi$, $\tf{A}\tf{X}_i \, \phi$ for $\neg \tf{E}\tf{X}_i \, \neg \phi$,
$\tf{A}\tf{F} \, \phi$ for
$\tf{A}[\true \, \tf{U} \, \phi]$, $\tf{E}\tf{F} \, \phi$ for $\tf{E}[\true
\, \tf{U} \, \phi]$, $\tf{E}\tf{G}
\, \phi$ for $\neg \tf{A}\tf{F} \, \neg \phi$, and  $\tf{A}\tf{G} \, \phi$
for $\neg \tf{E}\tf{F} \,
\neg \phi$.\\

\noindent{\em Semantics}: \vCTL formulas over a set of variables $V$
are interpreted over models of the form $M = (S, R, L)$, where $S$ is
a set of states and $R$ is a a total, multi-process, binary relation
$R = \cup_i R_i$ over $S$, composed of the transitions $R_i$ of 
each process $P_i$. $L$ is a labeling function that assigns to each
state $s \in S$ a valuation of all variables in $V$.  The value of a
term $t$ in a state $s \in S$ of $M$ is denoted as $\val^{(M,s)}[t]$,
and is defined inductively as in \secref{alpha}. A path in $M$ is a
sequence $\pi = (s_0, s_1, \ldots)$ of states such that $(s_j,
s_{j+1}) \in R$, for all $j \geq 0$. We denote the $j^{th}$ state in
$\pi$ as $\pi_j$. 


The satisfiability of a \vCTL formula in a state $s$ of $M$
can be defined as follows:

\begin{itemize}
\item $M,s \models G(t_1, \ldots, t_m)$ iff  $G(\val^{(M,s)}[t_1], \ldots,
\val^{(M,s)}[t_m]) = \true$. 
\item $M,s \models t_1 = t_2$ iff $\val^{(M,s)}[t_1] =
\val^{(M,s)}[t_2]$.
\item $M,s \models \neg \phi$ iff it is not the case that $M,s \models
\phi$.
\item $M,s \models \phi_1 \vee \phi_2$ iff $M,s \models \phi_1$ or
$M,s \models \phi_2$.
\item $M,s \models EX \, \phi$ iff for some $s_1$ such that  $(s,s_1)
\in R$, $M,s_1 \models \phi$. 
\item $M,s \models EX_i \, \phi$ iff for some $s_1$ such that
$(s,s_1) \in R_i$, $M,s_1 \models \phi$. 
\item $M,s \models A[\phi_1 \, U \, \phi_2]$ iff for all paths $\pi$
starting at $s$, $\exists j \, [M,\pi_j \models \phi_2 \text{ and }
\forall k \, (k < j \, \to \, M,\pi_k \models \phi_1)]$.
\item $M,s \models E[\phi_1 \, U \, \phi_2]$ iff there exists a path $\pi$
starting at $s$ such that $\exists j \, [M,\pi_j \models \phi_2 \text{ and }
\forall k \, (k < j \to M,\pi_k \models \phi_1)]$.\\
\end{itemize}


\noindent{\em Programs as Models}:
A program $P = (S,S^0,R)$ can be viewed as a model $M = (S,R,L)$, with
the same set of states and transitions as $P$, and the identity labeling
function $L$ that maps a state to itself. Given an $\vCTL$
specification $\phi$, we say $P \models \phi$ iff for each state $s
\in S^0$, $M,s \models \phi$. 


\section{Basic Algorithmic Framework}\label{sec:approach}
In this section, for ease of exposition, we assume a simpler program
model than the one described in \secref{prog}.  We restrict the number
of concurrent processes $k$ to $2$. We assume that {\em all} data
variables are initialized in the program
to specific values from their respective domains. We further assume
that all program variables, including control variables, are shared
variables.  We explain our basic algorithmic framework with these
assumptions, and later describe extensions to handle the general
program model in \secref{ext}.

Let us first review our problem definition.  Given a concurrent
program $P$, composed of unsynchronized processes $P_1$, $P_2$, and an
\vCTL specification $\phi_{spec}$ of their desired global concurrent
behaviour, we wish to automatically generate synchronized processes
$P^s_1$, $P^s_2$, such that the resulting concurrent program $P^s
\models \phi_{spec}$. If $P_1$, $P_2$ consist of atomic instructions,
we wish to obtain synchronization in the form of CCRs, with each
instruction enclosed in a CCR. In particular, the goal is to
synthesize the guard for each CCR, along with any necessary
(synchronization) assignments to be performed within the CCR.

We propose an automated framework to do this in several steps. 
\begin{enumerate}
\item Formulate an \vCTL formula $\phi_P$ to specify the semantics of the concurrent program $P$.
\item Construct a tableau $T_\phi$ for the formula $\phi$ given by
$\phi_P \wedge \phi_{spec}$. If $T_\phi$ is empty, declare specification as
inconsistent and halt. 
\item If $T_\phi$ is non-empty, extract a model $M$ for $\phi$ from it.
\item Decompose $M$ to obtain CCRs to synchronize each process.
\end{enumerate}

In what follows, we describe these steps in more detail. 

\subsection{Formulation of $\phi_P$}
\label{sec:high}

A reader familiar with the early synthesis work in
\cite{EmeCla82} will recall that the synthesis of a global model
requires a complete specification, which includes a temporal
description $\phi_P$ of the concurrency and operational semantics of
the unsynchronized concurrent program $P$, along with its desired
global behaviour $\phi_{spec}$. We propose to automatically infer an \vCTL
formula for $\phi_P$ to help mitigate the user's burden of
specification-writing. Let $Var = \{v_1, \ldots, v_h\}$ be the set of
data variables. \{$\phi_P$ is then formulated as the conjunction of
the following (classes of) properties: 

\begin{enumerate}
\item Initial condition:\\
$\val[loc_1] = l_1^0 \; \wedge \; \val[loc_2] = l_2^0 \; \wedge \;
\bigwedge_{v \in Var} \; \val[v] = \vinit$.
\item At any step, only one process can make a (local) move:\\
$\textsf{AG} \, \bigwedge_{j=1}^{j=n_1} \, ((\val[loc_1] = l_1^j) \,
\Rightarrow \, \textsf{AX}_2 \, (\val[loc_1] = l_1^j)) \quad \wedge \\
\textsf{AG} \, \bigwedge_{j=1}^{j=n_2} \, ((\val[loc_2] = l_2^j) \,
\Rightarrow \, \textsf{AX}_1 \, (\val[loc_2] = l_2^j))$.
\item Some process can always make a (local) move: \\
$\textsf{AG} (\textsf{EX}_1 \, \true \; \vee \; \textsf{EX}_2 \, \true)$.
\item A statement $l_i^r: \{v_{i1}, \ldots, v_{iq}\} \, {\bf
\assign} \, \{t_1, \ldots, t_q\}$ in $P_i$ is formulated as: \\
$\textsf{AG}(((\val[loc_i] = l_i^r) \, \wedge \; \bigwedge_{j=1}^{j=h}
\, \val[v_j] = \upsilon_j) \, \Rightarrow \,\\ \textsf{AX}_i \, ((\val[loc_i]
= l_i^{r+1}) \; \wedge \; \bigwedge_{j=1}^{j=q} \, \val[v_{ij}] = \val[t_j]
\; \wedge \; \bigwedge_{v_j  
\in Var \setminus \{v_{i1}, \ldots, v_{iq}\}} \, \val[v_j] = \upsilon_j))$.
\item A statement $l_i: {\tt if} \; (G) \; l_{if},\, l_{else}$ in $P_i$ is formulated as:\\
$\textsf{AG}(((\val[loc_i] = l_i) \, \wedge \, (\val[G] = \true)) \, \Rightarrow
\, \textsf{AX}_i \, (\val[loc_i] = l_{if})) \; \wedge \; \\
\textsf{AG}(((\val[loc_i] = l_i) \, \wedge \, (\val[G] = \false)) \, \Rightarrow
\, \textsf{AX}_i \, (\val[loc_i] = l_{else}))$.
\item A statement $l_i: {\tt goto} \; l$ in $P_i$ is formulated
as:\\
$\textsf{AG}((\val[loc_i] = l_i) \, \Rightarrow \, \textsf{AX}_i \, (\val[loc_i] = l))$
\end{enumerate}

\subsection{Construction of $T_\phi$}

We assume the ability to evaluate ${\bf L}$-atoms and ${\bf L}$-terms
over the set $V$ of program variables. Note that since we restrict ourselves
to a finite subset of the symbols in $\bL$, this is a reasonable
assumption. Let us further assume that the formula $\phi = \phi_P
\wedge \phi_{spec}$ is in a form in which only atoms appear negated. 

An {\em elementary}  formula of \vCTL is an atom, negation of an atom
or the formulas beginning with $\textsf{AX}_i$ or $\textsf{EX}_i$ (we do not explicitly
consider formulas beginning with $\textsf{AX}$ or $\textsf{EX}$ since
$\textsf{AX} \, \psi =
\bigwedge_i \, \textsf{AX}_i \, \psi$, and $\textsf{EX} \, \psi =
\bigvee_i \, \textsf{EX}_i \,
\psi$. All other formulas are nonelementary. Every nonelementary
formula is either a conjunctive formula $\alpha \equiv \alpha_1 \wedge
\alpha_2$ or a disjunctive formula $\beta \equiv \beta_1 \vee
\beta_2$. For example, $\psi_1 \wedge \psi_2$, $\tf{A}\tf{G}\, (\psi) =
\psi \, \wedge \, \tf{A}\tf{X}\tf{A}\tf{G} \, \psi$ are $\alpha$
formulas, and $\psi_1 \vee \psi_2$,  $\tf{A}\tf{F}\, (\psi) =
\psi \, \vee \, \tf{A}\tf{X}\tf{A}\tf{F} \, \psi$ are $\beta$ formulas. 

The tableau $T_\phi$ for the formula $\phi$ is a finite, rooted,
directed AND/OR graph with nodes labeled with formulas such that when
a node $B$ is viewed as a state in a suitable structure, $B \models
\psi$ for all formulas $\psi \in B$. The construction for $T_\phi$
is similar to the tableau-construction for propositional CTL in
\cite{EmeCla82}, while accounting for the presence of ${\bf L}$-atoms
over $V$ in the nodes of $T_\phi$. Besides composite ${\bf L}$-atoms
and \vCTL formulas, each node of $T_\phi$ is labeled with simple atoms
of the type $loc = l$ and $v = \upsilon$ identifying the values of the
control and data variables in each node. Two OR-nodes $B_1$ and $B_2$
are identified as being equivalent if $B_1$, $B_2$ are labeled with
the same simple atoms, and the conjunction of all the formulas in
$B_1$ is valid iff the conjunction of all the formulas in $B_2$ is
valid. Equivalence of AND-nodes can be similarly defined. We
briefly summarize the tableau construction first, before explaining
the individual steps in more detail.  

\begin{enumerate}
\item Initially, let the root node of $T_\phi$ be an OR-node labeled
with $\phi$.
\item If all nodes in $T_\phi$ have successors, go to the next step. 
Otherwise, pick a node $B$ without successors. Create appropriately 
labeled successors of $B$ such that: if $B$ is an OR-node, the formulas 
in $B$ are valid iff the formulas in some (AND-) successor node are 
valid, and if $B$ is an AND-node, the formulas in $B$ are valid iff 
the formulas in all (OR-) successor nodes are valid. Merge all 
equivalent AND-nodes and equivalent OR-nodes. Repeat this step.
\item Delete all {\em inconsistent} nodes in the tableau from the
previous step to obtain the final $T_{\phi}$.
\end{enumerate}

\noindent{\em Successors of OR-nodes}:
To construct the set of AND-node successors of an OR-node $B$, first
build a temporary tree with labeled nodes rooted at $B$, repeating the
following step until all leaf nodes are only labeled with elementary
formulas. For any leaf node $C$ labeled with a non-elementary formula
$\psi$: if $\psi$ is an $\alpha$ formula, add a single child node, 
labeled $C\setminus \{\psi\} \cup \{\alpha_1,\alpha_2\}$, to $C$, and if
$\psi$ is a $\beta$ formula, add two child nodes,  labeled $C\setminus
\{\psi\} \cup \{\beta_1\}$ and $C\setminus \{\psi\} \cup \{\beta_2\}$,
to $C$. Once the temporary tree is built, create an AND-node successor
$D$ for $B$, corresponding to each leaf node in the tree, labeled with the set
of all formulas appearing in the path to the leaf node from the root
of the tree. If there exists an atom of the form $v = t$ in $D$, where 
$t$ is an ${\bf L}$-term, and the valuation of $t$ in $D$ is $\upsilon$, 
replace the atom $v = t$ by the simple atom $v = \upsilon$. 
\\

\noindent{\em Successors of AND-nodes}: To construct the set of
OR-node successors of an AND-node $B$, create an OR-node labeled with
$\{\psi\}$ for each $EX_i \, \psi$ formula in $B$ and label the
transition to the OR-node with $i$. Furthermore, label each such
OR-node $D$ (with an $i$-labeled transition into $D$) with $\bigcup_j
\psi_j$ for each $AX_i \, \psi_j$ formula in $B$.  If there exists an
atom of the form $v = t$ in $D$, where $t$ is an ${\bf
L}$-term, and the valuation of $t$ in $D$ is $\upsilon$, replace the atom
$v = t$ by the simple atom $v = \upsilon$.  Note that the requirement that
some process can always move ensures that there will be some successor
for every AND-node. \\

\noindent{\em Deletion rules}: 
All nodes in the tableau that do not meet all criteria for a tableau for
$\phi$ are identified as inconsistent and deleted as follows:

\begin{enumerate}
\item Delete any node $B$ which is internally inconsistent, \ie, the
conjunction of all non-temporal elementary formulas in $B$ evaluates to
$\false$.
\item Delete any node all of whose original successors have been deleted. 
\item Delete any node $B$ such that $E[\psi_1 U \psi_2] \in B$, and
there does not exist some path to an AND-node $D$ from $B$ with $\psi_2
\in D$, and $\psi_1 \in C$ for all AND-nodes $C$ in the path. 
\item Delete any node $B$ such that $A[\psi_1 U \psi_2] \in B$, and
there does not exist a full sub-DAG \footnote{A full sub-DAG $T'$ is a
directed acyclic sub-graph of a tableau $T$, rooted at a node of $T$
such that all OR-nodes in $T'$ have exactly one (AND-node) successor
from $T$ in $T'$, and all AND-nodes in $T'$ either have no successors
in $T'$, or, have all their (OR-node) successors from $T$ in $T'$.} 
such that for all its frontier nodes $D$ , $\psi_2 \in D$ and for all 
its non-frontier nodes $C$, $\psi_1 \in C$.
\end{enumerate}

If the root node of the tableau is deleted, we halt and declare the
specification $\phi$ as inconsistent (unsatisfiable). If not, we proceed 
to the next step.

\subsection{Obtaining a model $M$ from $T_{\phi}$} 

A model $M$ is obtained by joining together model fragments rooted at
AND-nodes of $T_{\phi}$: each model fragment is a rooted DAG of
AND-nodes embeddable in $T_{\phi}$ such that all eventuality formulas
labeling the root node are fulfilled in the fragment. We do not
explain this step in more detail, as it is identical to the procedure
in \cite{EmeCla82} \footnote{There may be multiple models embedded in
$T_{\phi}$. In \cite{EmeCla82}, in order to construct model fragments,
whenever there are multiple sub-DAGs rooted at an OR-node $B$ that
fulfill the eventualities labeling $B$, one of minimal size is chosen,
where size of a sub-DAG is defined as the length of its longest path.
There are other valid criteria for choosing models, exploring which is
beyond the scope of this paper.}. After extracting $M$ from
$T_{\phi}$, we modify the labels of the states of $M$ by eliminating
all labels other than simple atoms, identifying the values of the
program variables in each state of $M$.  If there exist $n$ states
$s_1, \ldots, s_n$ with the exact same labels after this step, we
introduce an auxiliary variable $x$ with domain $\{0,1,2,\ldots,n\}$
to distinguish between the states: $x$ is assumed to be $0$ in all
states other than $s_1, \ldots, s_n$; for each $j \in \{1,\ldots,n\}$, we set
$x$ to $j$ in transitions into $s_j$, and set $x$ back to $0$ in
transitions out of $s_j$. This completes the model generation.  $M$ is
guaranteed to satisfy $\phi$ by construction. 

\subsection{Decomposition of $M$ into $P_1^s$ and
$P_2^s$}\label{sec:decomp} 

Recall that $P_1$ and $P_2$ are unsynchronized processes with atomic
instructions such as assignments, condition tests and gotos, and no
CCRs.  In this last step of our basic algorithmic framework, we
generate $P_1^s$ and $P_2^s$ consisting of CCRs, enclosing each
atomic instruction of $P_1$ and $P_2$. 

Without loss of generality, consider location $l_1$ in $P_1$. The
guard for the CCR for $inst(l_1)$ in $P^s_1$ corresponds to all states
in $M$ in which $inst(l_1)$ is {\em enabled}, \ie, states in which
$P_1$ is at location $l_1$ and from which there exists a $P_1$
transition. To be precise, $inst(l_1)$ is enabled in state $s$ in $M$
iff there exists a transition $(s,s') \in R$ such that $\val^s[loc_1]
= l_1$, $\val^{s'}[loc_2] = l'_1$ with $l'_1$ being a valid next
location for $P_1$, and, $\val^s[loc_2] = \val^{s'}[loc_2]$. The guard
$G_{s}$ corresponding to such a state $s$ is the valuation of all
program variables other than $loc_1$ in state $s$. Thus, if
$\val^s[loc_2] = l_2$ and for all $v_j \in Var = \{v_1,\ldots,v_h\}$,
$\val^s[v_j] = \upsilon_j$, then $G_{s}$ is given by $(loc_2 = l_2) \,
\wedge \; \bigwedge_{j=1}^{j=h} \, v_j = \upsilon_j$.

If $M$ does not contain an auxiliary variable, then the CCR for
$inst(l_1)$ in $P^s_1$ is simply $G_{1,1} \to inst(l_1)$, where
$G_{1,1}$ is the disjunction of guards $G_s$ corresponding to all
states $s$ in $M$ in which $inst(l_1)$ is enabled.  However, if $M$
contains an auxiliary variable $x$ (with domain $\{0,1,2,\ldots,n\}$),
then one may also need to perform updates to $x$ within the CCR
instruction block. In particular, if $inst(l_1)$ is enabled on state
$s$ in $M$, transition $(s,s')$ in $M$ is a $P_1$ transition, and if there is
an assignment $x \, \assign \, j$ for some $j \in \{0,\ldots,n\}$ along
transition $(s,s')$, then besides $inst(l_1)$, the instruction block
of the CCR for $inst(l_1)$ in $P^s_1$ includes instructions in our
programming language corresponding to: ${\tt if} \; G_s \; x \, \assign \, j$.  

The synchronized process $P_1^s$ (and similarly $P_2^s$) can be
generated by inserting a similarly generated CCR at each location in
$P_1$ (and $P_2$).  The modified concurrent program $P_s$ is given by
$P_s::[{\tt declaration}] \, [P_1^s {\tt \parallel} P_2^s]$, where the
declaration includes auxiliary variable $x$ with domain
$\{0,1,2,\ldots,n\}$ if $M$ contains $x$ with domain $\{0,1,2,\ldots,n\}$. 

\subsection{Correctness and Complexity}

The following theorems assert the correctness of our basic algorithmic
framework for synthesizing synchronization for unsynchronized
processes $P_1$, $P_2$, as defined in \secref{prog}, with the
restriction that all program variables are shared variables that are
initialized to specific values. 

\begin{thm}
Given unsynchronized processes $P_1$, $P_2$ and an $\vCTL$ formula
$\phi_{spec}$, if our basic algorithm generates $P^s$, then $P^s \models
\phi_{spec}$.
\end{thm}

\begin{thm}
Given unsynchronized processes $P_1$, $P_2$, and an $\vCTL$ 
formula $\phi_{spec}$, if the temporal specification $\phi =
\phi_{spec} \wedge
\phi_P$ is consistent as a whole, then our method constructs $P^s$ 
such that $P^s \models \phi_{spec}$.
\end{thm}

The complexity of our method is exponential in the size of $\phi$,
\ie, exponential in the size of $\phi_{spec}$ and the number of program
variables $V$.

\section{Extensions}\label{sec:ext}
In this section, we demonstrate the adaptability of our basic
algorithmic framework by considering more general program models. In
particular, we discuss extensions for synthesizing correct
synchronization in the presence of
uninitialized variables and local variables. Furthermore, we extend
our framework to programming languages with locks and {\tt
wait}/{\tt signal} over condition variables by presenting an automatic
compilation of CCRs into synchronization code based on these
lower-level synchronization primitives. We conclude with an extension
of the framework to multiple processes. 

\subsection{Uninitialized Variables}

In \secref{approach}, we assumed that all data variables are
initialized to specific values over their domains. This assumption may
not be satisfied in general as it disallows any kind of user or
environment input to a concurrent program. In the program model
presented in \secref{prelim}, only some (or even none) of the data
variables may be initialized to specific values within the program.
This is a more realistic setting, which allows a user or environment
to choose the initial values of the remaining data variables. In this
subsection, we present a simple, brute-force extension of our basic
algorithm for synthesizing synchronization in the presence of
uninitialized variables. 

The formula $\phi_P$, expressing the concurrency and operational
semantics of $P$, remains the same, except for the initial condition.
Instead of a single initial state, the initial condition in $\phi_P$
specifies the set of all possible initial states, with the control and
initialized data variables set to their initial values, and the
remaining data variables ranging over all possible values in their
respective domains. Let us denote by $Var_{inp}$ this remaining set of
data variables, that are, essentially, inputs to the program $P$. The
set of program-initialized data variables is then $Var \setminus
Var_{inp}$. The initial condition in $\phi_P$ is expressed as: 

$\bigwedge_i \, \val[loc_i] = l_i^0 \; \wedge \; 
\bigwedge_{v \in Var \setminus Var_{inp}} \, (v = \vinit) \; \wedge
\; \bigwedge_{v \in Var_{inp}} \, \bigvee_{\upsilon_j \in D_v} (v = \upsilon_{j})$, 

\noindent where $D_v$ is the domain of $v$.

The root node of the tableau $T_{\phi}$ is now an AND-node with
multiple OR-node successors, each corresponding to a particular
valuation ${\bf \upsilon}$ of all the data variables (the values of
the control variable and initialized data variables are the same in
any such valuation). Each such OR-node yields a model $M_{{\bf
\upsilon}}$ for the formula $\phi$, and a corresponding decomposition
of $M_{{\bf \upsilon}}$ into synchronized processes $P_{1_{{\bf
\upsilon}}}^s$ and $P_{2_{{\bf \upsilon}}}^s$.  

To generate synchronized processes $P_1^s$ and $P_2^s$ such that for
all possible initial valuations ${\bf \upsilon}$ of the data
variables, $P^s \models \phi_{spec}$, we propose to {\em unify} the
CCRs corresponding to each valuation ${\bf \upsilon}$ as follows: 

\begin{enumerate}
\item Introduce a new variable $v0$ for every input data variable
$v$ in $Var_{inp}$. Declare $v0$ as a variable with the same domain as $v$. 
Assign $v0$ the input value of $v$. 
\item Replace every CCR guard $G$ in $P_{i_{{\bf \upsilon}}}^s$ with the
guard $G_{\bf \upsilon}$ given by $\bigwedge_{v \in Var_{inp}} (v0 = {\bf
\upsilon}_v) \,
\wedge \, G$, where the valuation of $v$ in ${\bf \upsilon}$ is ${\bf
\upsilon}_v$. Similarly, update every conditional guard accompanying
an auxiliary variable assignment within a CCR instruction block in
$P_{i_{{\bf \upsilon}}}^s$.
\item The unified guard for each CCR in $P_1^s$ and $P_2^s$ is given 
by the disjunction of the corresponding guards $G_{\bf \upsilon}$ 
in all $P_{1_{{\bf \upsilon}}}^s$ and $P_{2_{{\bf \upsilon}}}^s$. 
The unified conditional guards for auxiliary variable updates in the CCR
instruction blocks are computed similarly.  
\end{enumerate}

Note that the unified guards inferred above, as well as in
\secref{decomp}, may not in general be {\em pleasant}. However, since
each guard  is expected to an ${\bf L}$-term over a finite set of
variable, function and predicate symbols with known interpretations,
it is possible to obtain a simplified ${\bf L}$-term with the same
value as the guard. This translation is beyond the scope of this
paper, but we refer the reader to \cite{KMPS12} for a similar
approach.

\subsection{Local Variables}

Another assumption in \secref{approach} was that all program
variables, including control variables, were shared variables. Since
one typically associates a cost with each shared variable access, it
is impractical to expect all program variables to be shared variables.
This is especially true of control variables, which are generally
never declared explicitly or accessed in programs. Thus, the guards
inferred in \secref{decomp}, ranging over locations of the other
process, are somewhat irregular. Indeed, any guard for a process
$P_i$ must only be defined over the data variables $Var_i$ accessible by
$P_i$. In what follows, we discuss various solutions to address this
issue.  

Let us assume that we have a model $M=(S,R,L)$ for $\phi$, with states
labeled by the valuations of the control variables $Loc$, the shared
data variables $X$, the local data variables $Y = \bigcup_i Y_i$, and
possibly an auxiliary variable $x$. For the purpose of this
subsection, let $x$ be included in the set $X$. We first check if the
set of states $S$ of $M$ has the property that for any two states
$s_1$, $s_2$ in $S$: $[\bigwedge_{loc \in Loc} \, \val^{s_1}[loc] =
\val^{s_2}[loc] \; \wedge \; \bigwedge_{y \in Y} \val^{s_1}[y] =
\val^{s_2}[y]] \; \Leftrightarrow \; \bigwedge_{x \in X} \val^{s_1}[x]
= \val^{s_2}[x]$. If this
is true, then each state $s \in S$ is uniquely identified by its
valuation of the shared data variables $X$. We can then simply factor
out guards from $M$ for each process that only range over $X$, without
missing out on any permitted behaviour in $M$.  If this is not true,
we can perform other similar checks. For instance, we can check if for
a particular $i$: any two states in $S$ match in their valuations of
the variables $\{loc_i\} \cup Y_i \cup X$ iff they match in their
valuations of the other program variables. If this is true, then the
process $P_i$ can distinguish between states in $S$ by the
valuations of its variables $Var_i \cup \{loc_i\}$. Thus, we can
infer guards for $P_i$, that are equivalent to the guards inferred in
\secref{decomp}, but only range over $Var_i$. 

In general, however, there will be states $s_1$, $s_2$ in $S$ which
cannot be distinguished by the valuations of a particular process's,
or worse, by any process's variables. This general situation presents
us with a trade-off between synchronization cost and concurrency: we
can introduce additional shared variables to distinguish between such
states, thereby increasing the synchronization cost and allowing more
behaviours of $M$ to be preserved in $P^s$, or, we can resign to {\em
limited observability} \cite{VYY09} of global states, resulting in
lower synchronization cost and fewer permitted behaviours of $M$. In
particular, for the latter case, we implement a safe subset of the
behaviours of $M$ by inferring synchronization guards corresponding to
the negation of variable valuations (states) that are not
present in $M$. Since a global state $u \not \in M$  may be
indistinguishable over some $Var_i$ from a state $s \in M$, when
eliminating behaviours rooted at $u$, we also eliminate all (good)
behaviours of $M$, rooted at $s$. We refer the reader to \cite{VYY09}
for a detailed treatment of this trade-off.

\subsection{Synchronization using Locks and Condition
Variables}

While CCRs provide an elegant high-level synchronization solution,
many programming languages prefer and only provide lower-level
synchronization primitives such as locks for mutual exclusion,
and  {\tt wait}/{\tt signal} over condition variables for condition
synchronization. In what follows, we present an automatic compilation of
the CCRs inferred in \secref{decomp} for $P^s_1$, $P^s_2$ into both
coarse-grained and fine-grained synchronization code based on these
lower-level primitives. The resulting processes are denoted as
$P^c_1$, $P^c_2$ (coarse-grained) and $P^f_1$, $P^f_2$ (fine-grained). 

In both cases, we declare locks and 
conditions variables for synchronization. For the program $P^c$, which
has a coarser level of lock granularity, we declare a single lock $l$
for controlling access to shared variables and condition variables.
For the program $P^f_1 \parallel P^f_2$ with a finer level of lock
granularity, we declare separate locks $l_v$, $l_x$ for
controlling access to each shared data variable $v \in X$ and the
shared auxiliary variable $x$, respectively. We further define a
separate lock $l_{cv_{1,i}}$, $l_{cv_{2,j}}$ for each
condition variable $cv_{1,i}$, $cv_{2,j}$ to allow simultaneous
processing of different condition variables.

\begin{figure}[t]
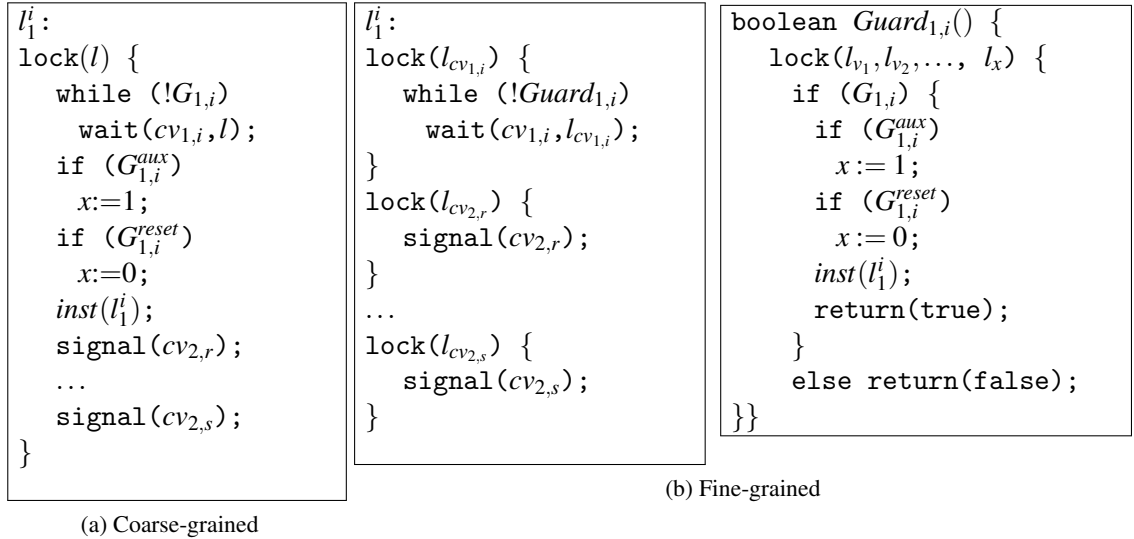

\centering
\subfloat[Coarse-grained]{
\label{fig:p1coarse}
\fbox{
\begin{minipage}[t]{0.27\linewidth}
{\tt
$l_1^i$:\\
$\lk{l}$ \{\\
\hspace*{3mm} while ($!G_{1,i}$)\\
\hspace*{6mm} \wait{$cv_{1,i}$}{$l$};\\
\hspace*{3mm} if ($G^{aux}_{1,i}$) \\
\hspace*{6mm} $x \assign 1$;\\
\hspace*{3mm} if ($G^{reset}_{1,i}$) \\
\hspace*{6mm} $x \assign 0$;\\
\hspace*{3mm} $inst(l_1^i)$;\\
\hspace*{3mm} \sig{$cv_{2,r}$};\\
\hspace*{3mm} $\ldots$\\
\hspace*{3mm} \sig{$cv_{2,s}$};\\
\}\\
}
\end{minipage}}}
\subfloat[Fine-grained]{
\label{fig:p1fine}
\fbox{
\begin{minipage}[t]{0.28\linewidth}
{\tt
$l_1^i$:\\
\lk{$l_{cv_{1,i}}$} \{\\
\hspace*{3mm} while ($!Guard_{1,i}$)\\
\hspace*{6mm} \wait{$cv_{1,i}$}{$l_{cv_{1,i}}$};\\
\}\\
\lk{$l_{cv_{2,r}}$} \{\\
\hspace*{3mm} \sig{$cv_{2,r}$};\\
\}\\
$\ldots$\\
\lk{$l_{cv_{2,s}}$} \{\\
\hspace*{3mm} \sig{$cv_{2,s}$};\\
\}\\
}
\end{minipage}}
\hspace{0mm}
\fbox{
\begin{minipage}[t]{0.33\linewidth}
{\tt
boolean $Guard_{1,i}()$ \{\\
\hspace*{3mm} lock($l_{v_1}, l_{v_2}, \ldots$, $l_x$) \{\\
\hspace*{6mm} if ($G_{1,i}$) \{\\
\hspace*{9mm} if ($G^{aux}_{1,i}$) \\
\hspace*{12mm} $x \; \assign \; 1$;\\
\hspace*{9mm} if ($G^{reset}_{1,i}$) \\
\hspace*{12mm} $x \; \assign \; 0$;\\
\hspace*{9mm} $inst(l_1^i)$;\\
\hspace*{9mm} return(true);\\
\hspace*{6mm} \}\\
\hspace*{6mm} else return(false);\\
\}\}
}
\end{minipage}}
}
\caption{Coarse and fine-grained synchronization code corresponding to
an example CCR at location $l_1^i$ of $P_1$. Guards $G^{aux}_{1,i}$, 
$G^{reset}_{1,i}$ above corresponds to all states in $M$ on which
$inst(l_1^i)$ is enabled, and there's an assignment $x \assign 1$, $x
\assign 0$, respectively, along a $P_1$ transition out of the states.}
\label{fig:p1sync}
\end{figure}

We refer the reader to \figref{p1coarse} for an example of
coarse-grained synchronization code corresponding to the CCR at
location $l_1^i$ of $P_1$. Note that, for ease of presentation, we
have used conventional pseudocode, instead of our programming
language. Further note that we find it convenient to express locks, as
{\tt lock(l)}$\{ \ldots \}$ (in a  manner similar to Java's {\tt
synchronized} keyword), wherein $l$ is a lock variable, `$\{$' denotes
{\em lock acquisition} and `$\}$' denotes {\em lock release}.  This
simple implementation involves acquiring the lock $l$ and checking if
the overall guard $G_{1,i}$ for executing $inst(l_1^i)$ is enabled.
While the guard is $\false$, $P^c_1$ {\em waits} for it to change to
$\true$.  This is implemented by associating a condition variable
$cv_{1,i}$ with the overall guard $G_{1,i}$: $P^c_1$ releases the lock
$l$ and waits till $P^c_2$ {\em signals} it that $G_{1,i}$ could be
$\true$; $P^c_1$ then reacquires the lock and rechecks the guard.  If
the overall guard is $\true$, $P_c^1$ enters the instruction block of
the CCR and executes the instructions while holding the lock $l$.
Finally, $P^c_1$ sends a notification signal corresponding to every
guard (\ie condition variable) of $P^c_2$ which may be changed to
$\true$ by $P^c_1$'s shared variables updates, and releases the lock.

While fine-grained locking can typically be achieved by careful
definition and nesting of multiple locks, one needs to be especially
cautious in the presence of condition variables for various reasons.
For instance, upon execution of \wait{$c$}{$l$} in a nested locking
scheme, a process only releases the lock $l$ before going to sleep,
while still holding all outer locks. This can potentially lead to a
deadlock.  The fine-grained synchronization code in $P^f_1$, shown in
\figref{p1fine}, circumvents these issues by utilizing a separate
subroutine to evaluate the overall guard $G_{1,i}$. In this
subroutine, $P^f_1$ first acquires all necessary locks, corresponding
to all shared variables accessed in the subroutine.  These locks are
acquired in a strictly nested fashion and in a predecided fixed order
to prevent deadlocks.  We use {\tt lock($l_1,l_2,\ldots$)}$\{ \ldots
\}$ to denote the nested locks \lk{$l_1$}\{ \lk{$l_2$}\{ $\ldots$\}\},
with $l_1$ being the outermost lock variable. The subroutine then
evaluates $G_{1,i}$ and returns its value to the main body of $P^f_1$.
If found $\true$, the subroutine also executes the instruction block
of the CCR.  The synchronization code in the main body of $P^f_1$
acquires the relevant lock $l_{cv_{1,i}}$ and calls its
guard-computing subroutine within a {\tt while} loop till it returns
$\true$, after which it releases the lock.  If the subroutine returns
$\false$, the process releases $l_{cv_{1,i}}$ and waits on the
associated condition variable $cv_{1,i}$ .  Each notification signal
for a condition variable, on which the other process may be waiting,
is sent out by acquiring the corresponding lock.

We emphasize certain optimizations implemented in our compilations
that potentially improve the performance of the synthesized concurrent
program: (a) declaration of condition variables only when necessary,
and (b) sending notification signals only when some guard in the other
process may have changed. We refer the reader to \cite{EmeSam11} for
more details of this compilation. 

\subsection{Multiple ($k > 2$) Processes}

Our basic algorithmic framework can be extended in a straight-forward
manner to the synthesis of synchronization for concurrent programs
with an arbitrary (but fixed) number $k$ of processes. But since this
involves building a global model $M$, with size exponential in $k$, it
exhibits a state explosion problem. There has, however, been work
\cite{AttEme89,Attie99} on improving the scalability of the approach
by avoiding building the entire global model, and instead composing
interacting process pairs to generate synchronized processes. Hence,
for $k > 2$ processes, we can adapt the more scalable synthesis
algorithms to the synthesis of ${\bf L}CTL$ formulas.

The compilation of CCRs into coarse-grained and fine-grained
synchronization code can be extended in a straight-forward manner to
$k > 2$ processes. We emphasize that this compilation acts on
individual processes directly, without construction or manipulation of
the global model, and hence circumvents the state-explosion problem
for arbitrary $k$.

\section{Discussion}\label{sec:disc}

\noindent{\em Related work}: Early work on synthesis of
synchronization for shared-memory concurrent programs from temporal
specifications \cite{EmeCla82} utilized a tableau-based decision
procedure for extracting synchronization skeletons from unsynchronized
process skeletons.  While the core technique has great potential, the
original work had little practical impact due to its remoteness from
realistic concurrent programs and programming languages. The limited
modeling of shared-memory concurrency in this work did not include
local and shared data variables, and hence, could not support semantic
specifications over the values of program variables. There was no
explicit treatment of process skeletons with branching, observability
of program counters or local variables, and no attempt to synthesize
synchronization based on lower-level synchronization primitives.

More recently, practically viable synthesis of synchronization has
been proposed for both finite-state \cite{VYY09} and infinite-state
concurrent programs \cite{VYY10}. However, in both \cite{VYY09},
\cite{VYY10}, the authors only handle safety specifications; in fact,
it can be shown that synthesis methods that rely on pruning a global
product graph (\cite{JanMyc06,VYY09,VYY10}) cannot, in general, work
for liveness.  Moreover, these papers do not support any kind of
external environment; in particular, these papers do not account for
different (environment-enabled) initializations of the program
variables.  Finally, similar to \cite{EmeCla82}, these papers only
synthesize high-level synchronization in the form in CCRs \cite{VYY09}
and atomic sections \cite{VYY10}, and do not attempt to synthesize
synchronization based on lower-level synchronization primitives
available in commonly used programming languages.

On the other end of the spectrum, there has been some important work
on automatic synthesis of lower-level synchronization, in the form of
memory fences, for concurrent programs running on relaxed memory models
\cite{LNPVY12,KVY10}.  There has also been work on mapping high-level
synchronization into lower-level synchronization
\cite{DDHM01,YavBul02} - these papers do not treat liveness
properties, are not fully algorithmic, and are verification-driven.
Among papers that address refinement of locking granularity, are
\cite{AttEme96}, which translates guarded commands, into
synchronization based on atomic reads and atomic writes, and papers on
compiler-based lock inference for atomic sections (\cite{EFJM07},
\cite{CCG08} \etc). The lock-inference papers \cite{EFJM07},
\cite{CCG08} rely on the availability of high-level synchronization in
the form of atomic sections, and do not, in general, support condition
synchronization.  Sketching \cite{SRBE05}, a search-based program
synthesis technique, is a verification-driven approach, which can be
used to synthesize optimized implementations of synchronization
primitives, \eg barriers, from partial program sketches.\\

\noindent{\em A note on reactive systems}: 
A shared-memory concurrent program can also be viewed as a {\em reactive
system}.  A {\em reactive system} \cite{HarPnu85, Pnueli85} is described as
one that maintains an ongoing interaction with an external environment
or within its internal concurrent modules. Such systems cannot be
adequately described by relational specifications over initial and
final states - this distinguishes them from transformational or
relational programs.  An adequate description of a reactive system
must refer to its ongoing desired behaviour, throughout its (possibly
non-terminating) activity - temporal logic \cite{Pnueli77} has been recognized
as convenient for this purpose. 

A reactive system may be terminating or not, sequential or concurrent,
and implemented on a monolithic or distributed architecture.  A
reactive system can also be open or closed \cite{PnuRos89a, PnuRos90}.
This has been a somewhat overlooked dichotomy in recent years. We have
observed that it is not uncommon to view reactive systems exclusively
as open systems; this is especially true in the context of synthesis. 
 While the first algorithms on synthesis of concurrent
programs \cite{EmeCla82, ManWol84, AttEme89} were proposed for closed
reactive systems, the foundational work in \cite{PnuRos89a, PnuRos90}
set the stage for an extensive body of impressive results on synthesis
of open reactive systems (see \cite{Thomas09} for a survey). 

We contend that the relatively simpler problem of synthesis of closed
reactive systems is an important problem in its own right. This is
especially true in the context of shared-memory concurrent programs,
where it is sometimes sufficient and desirable to model 
programs as closed systems and force the component processes to
cooperate with each other for achieving a common goal. If one must
model an external environment, it is also often sufficient to model
the environment in a restricted manner (as in this paper) or
optimistically assume a helpful environment (see \cite{AlfHen01}). \\

\noindent{\em Concluding Remarks}: 
In this paper, we have presented a general tableau-based framework for
the synthesis of synchronization for shared memory concurrent
programs. While we have identified and explored initial solutions for
issues such as environment-initialized variables, limited
observability of local variables, pleasantness of guards, much work
remains to be done. We also wish to extend the basic program model to
handle nondeterministic programs, infinite-state programs as well as
dynamic allocation of threads.  Finally, we want to investigate
techniques to reduce the overall complexity of the method. \\

\noindent{\em Acknowledgements}: 
The author wishes to thank Jyotirmoy Deshmukh for many insightful 
discussions during the course of writing this paper, and an anonymous
reviewer for pointing out interesting future research directions. 

\bibliographystyle{eptcs}
\tiny{\bibliography{refs}}


\end{document}